\documentclass[prl,twocolumn,showpacs]{revtex4}
\usepackage{amsmath}
\usepackage{amsfonts}

\begin{document}
\title{Efficient Classical Simulation of Continuous Variable
  Quantum Information Processes}
\author{Stephen D.\ Bartlett}
\author{Barry C.\ Sanders}
\affiliation{Department of Physics and Centre for Advanced Computing
  -- Algorithms and Cryptography, \\ Macquarie University, Sydney, New
  South Wales 2109, Australia}
\author{Samuel L.\ Braunstein}
\author{Kae Nemoto}
\affiliation{Informatics, Bangor University, Bangor LL57 1UT, UK}
\date{November 22, 2001}

\begin{abstract}
  We obtain sufficient conditions for the efficient simulation of a
  continuous variable quantum algorithm or process on a classical
  computer.  The resulting theorem is an extension of the
  Gottesman-Knill theorem to continuous variable quantum information.
  For a collection of harmonic oscillators, any quantum process that
  begins with unentangled Gaussian states, performs only transformations
  generated by Hamiltonians that are quadratic in the canonical
  operators, and involves only measurements of canonical operators
  (including finite losses) and suitable operations conditioned on these
  measurements can be simulated efficiently on a classical computer.
\end{abstract}
\maketitle

Quantum mechanics allows for information processing that could not be
performed classically.  In particular, it may be possible to perform
an algorithm efficiently on a quantum computer that cannot be
performed efficiently on a classical one.  Significant effort is now
underway to construct quantum algorithms and processes that yield such
a speedup.  The Gottesman-Knill (GK) theorem~\cite{Got99} for
discrete-variable (qubit) quantum information provides a valuable tool
for assessing the classical complexity of a given process.
Essentially, it states that any quantum algorithm that initiates in
the computational basis and employs only a restricted class of gates
(Hadamard, phase, CNOT, and Pauli gates), along with projective
measurements in the computational basis, can be efficiently simulated
on a classical computer. (For a precise formulation and proof of this
remarkable theorem, see~\cite{Nie00}, page 464.)  The GK theorem
reveals that a large class of quantum algorithms do not provide a
speedup over classical processes.  In fact, recent work has placed
even stronger constraints on the potential speedup of fermionic
quantum computers~\cite{Kni01}.

In addition to the successes of qubit-based algorithms, quantum
information over continuous variables (CV) has yielded many exciting
advances, both theoretically and experimentally, in fields such as
quantum teleportation~\cite{Bra98,Fur98}, quantum
cryptography~\cite{Ral00,Hil00,Rei00,Got01a}, and potentially quantum
computation~\cite{Llo99}.  CV algorithms could also perform
computational tasks more efficiently than is possible classically.  To
assess the computational complexity of these tasks, it is necessary to
develop an extension of the GK theorem: what continuous variable
processes can be efficiently simulated on a classical computer?  As a
CV quantum information process involves coupled canonical systems,
this question of efficient classical simulation is related to asking
under what conditions a quantum mechanical system can be modeled by a
classical one.  As noted by Feynman~\cite{Fey86}, a key advantage of a
quantum computer is its ability to simulate quantum systems that
cannot be efficiently simulated classically.

The issue of efficient classical simulation of a CV process is more
involved than for the discrete case.  One notable difference is that
the quantum states and the unitary transformations involved are
described by real-valued (as opposed to integer-valued) parameters,
and these parameters must be described on a discrete classical
computer with some assumption of error or limited precision.  Also,
the states used in CV experiments are approximations to the idealized
computational basis. These basis states are infinitely squeezed states
whereas any experimental implementation will involve finitely squeezed
states thus deviating from their idealized form~\cite{Llo99}.  A good
classical simulation must be robust against such deviations.
Measurements are part of the quantum computation and, even in the
computational basis, are subject to experimental constraints (such as
photodetection efficiency).  Classical simulation must also
incorporate these measurements.

Despite these complications, we prove in the following an extension of
the GK theorem for continuous variables; \textit{i.e.}, we present a
set of sufficient conditions for a CV quantum information process
which, if satisfied, ensure that it can be efficiently simulated on a
classical computer.  To prove this theorem, we employ the techniques
of stabilizers~\cite{Nie00} that are used for qubits.  Using the
stabilizer formalism, it is often possible to simulate a quantum
information process by following the evolution of a set of operators,
the Pauli operators, rather than the evolution of quantum states.  For
CV processes, we show that it is more natural to analyze stabilizers
in terms of the \emph{algebras} (\textit{i.e.}, Hamiltonians) that
generate them, rather than the groups themselves.  We define analogs
of the Pauli and Clifford algebras and groups for CV and construct
sets of gates (as unitary transformations) that can efficiently
simulate any arbitrary transformation in these groups.  Any algorithm
or process constructed out of these Clifford group transformations can
be efficiently modeled by following the evolution of the Pauli
operators rather than the states of the system.

The standard Pauli group $\mathcal{G}_n$ for CV quantum computation on
$n$ coupled oscillator systems is the Heisenberg-Weyl group HW($n$),
which consists of phase-space displacement operators for the $n$
oscillators.  Unlike the discrete Pauli group for qubits, the group
HW($n$) is a continuous (Lie) group, and can therefore only be
generated by a set of continuously-parameterized operators.  The
algebra hw($n$) that generates this group is spanned by the $2n$
canonical operators $\hat{q}_i$, $\hat{p}_i$, $i=1,\ldots,n$, along
with the identity operator $\hat{I}$, satisfying the commutation
relations $[\hat{q}_i,\hat{p}_j] = i\hbar \delta_{ij} \hat{I}$.  For a
single oscillator, the $n=1$ algebra is spanned by the canonical
operators $\{ \hat{q}, \hat{p}, \hat{I} \}$ which generate the single
oscillator Pauli operators
\begin{equation}
  \label{eq:PauliOperators}
  X(q) = e^{-\frac{i}{\hbar}q \hat{p}} \, , \quad
  Z(p) = e^{\frac{i}{\hbar}p \hat{q}} \, ,
\end{equation}
with $q,p \in \mathbb{R}$.  The Pauli operator $X(q)$ is a
position-translation operator (translating by an amount $q$), whereas
$Z(p)$ is a momentum boost operator (kicking the momentum by an
amount $p$).  These operators are non-commutative and obey the
identity
\begin{equation}
  \label{eq:PauliOperatorCommutation}
  X(q) Z(p) = e^{-\frac{i}{\hbar}qp} Z(p) X(q) \, .
\end{equation}

On the computational basis of position eigenstates $\{ |s\rangle; s
\in \mathbb{R}\}$ \cite{Llo99,Bra98b,Bra98c}, the Pauli operators act as
\begin{equation}
  \label{eq:ActionPauliOnCompBasis}
  X(q)|s\rangle = |s + q\rangle \, , \quad
  Z(p)|s\rangle = \exp(\frac{i}{\hbar}ps)|s\rangle \, .
\end{equation}
Note that it is conventional to use highly squeezed states to approximate 
position eigenstates;  these states satisfy the orthogonality relation 
$\langle s|s'\rangle = \delta(s-s')$ in the limit of infinite squeezing.

The Pauli operators for one system can be used to construct a set of
Pauli operators $\{ X_i(q_i), Z_i(p_i); i=1,\ldots,n \}$ for $n$
systems (where each operator labeled by $i$ acts as the identity on
all other systems $j \neq i$).  This set generates the Pauli group
$\mathcal{G}_n$.  Note that the Pauli group is only a subgroup of all
possible unitary transformations.  It is not possible to construct an
arbitrary unitary transformation using only the Pauli operators $X(q)$
and $Z(p)$; the Pauli group only describes transformations generated
by Hamiltonians that are linear in the canonical variables.

For issues of classical simulation, we will be interested in
transformations that lie in the \emph{Clifford group}.  The Clifford
group $N(\mathcal{G}_n)$ is the group of transformations, acting by
conjugation, that preserves the Pauli group $\mathcal{G}_n$;
\textit{i.e.}, it is the normalizer of the Pauli group in the
(infinite-dimensional) group of all unitary transformations.

\textbf{Theorem 1:} \textit{The Clifford group $N(\mathcal{G}_n)$ for
  continuous variables is the semidirect product group {\rm
    [HW($n$)]Sp($2n,\mathbb{R}$)}, consisting of all phase-space
  translations along with all one-mode and two-mode squeezing
  transformations.  This group is generated by inhomogeneous quadratic
  polynomials in the canonical operators.}

\textit{Proof:} The most straightforward method to identify the
Clifford group will be to identify its algebra.  The Clifford algebra
consists of all Hamiltonian operators $\hat{H}_{\text{c}}$ satisfying
$[\hat{H}_{\text{hw}},\hat{H}_{\text{c}}] \in \text{hw}(n)$ for all
$\hat{H}_{\text{hw}} \in$ hw($n$).  This algebra must obviously
include the algebra hw($n$), and thus hw($n$) is a subalgebra of the
Clifford algebra.  In addition, this algebra includes all homogeneous
quadratic polynomials in the canonical operators $\{ \hat{q}_i,\,
\hat{p}_i;\, i=1,\ldots,n \}$.  This algebra of quadratics consists of
Hamiltonians that generate one-mode squeezing transformations [for
example, the Hamiltonian $\hat{H}_S = \tfrac{1}{2}(\hat{q}\hat{p} +
\hat{p}\hat{q})$], and also interaction Hamiltonians that generate
two-mode squeezing transformations (for example, the interaction
Hamiltonian $\hat{H}_{\text{int}} = \hat{q}_1 \otimes \hat{p}_2$).
The algebra of homogeneous quadratic polynomials in the canonical
operators is known as the linear symplectic algebra
sp($2n,\mathbb{R}$).

Together, the algebras hw($n$) and sp($2n,\mathbb{R}$) form a larger
algebra, consisting of \emph{inhomogeneous} quadratic Hamiltonians in
the canonical operators $\{ \hat{q}_i, \hat{p}_i; i=1,\ldots,n \}$.
This algebra is the semidirect sum algebra
[hw($n$)]sp($2n,\mathbb{R}$), with hw($n$) as an ideal.  The group
generated by this algebra is the semidirect product group
[HW($n$)]Sp($2n,\mathbb{R}$).  This group includes phase-space
displacements (the Pauli group), as well as the squeezing
transformations (both single- and two-mode) of quantum
optics~\cite{Wal94}.  \hfill\textit{(QED)}

In order to describe a quantum information process as a circuit, it is
necessary to find a set of transformations (gates) that generate the
Clifford group; these gates will serve as building blocks for
arbitrary Clifford group transformations.  Following the derivation by
Gottesman \emph{et al}~\cite{Got01b}, a set of gates will
be defined in terms of the elements of the Clifford algebra
(\textit{i.e.}, the Hamiltonians) that generate the transformations.

The SUM gate is the CV analog of the CNOT gate and provides the basic
interaction gate for two oscillator systems $1$ and $2$; it is defined
as
\begin{equation}
  \label{eq:DefSUM}
  \text{SUM} = \exp \Bigl( -\frac{i}{\hbar} \hat{q}_1 \otimes
  \hat{p}_2 \Bigr)\, .
\end{equation}
This gate is an interaction gate operation on the Pauli group
$\mathcal{G}_2$ for two systems.  Referring to the definition
(\ref{eq:PauliOperators}) for the Pauli operators for a single system,
the action of this gate on the $\mathcal{G}_2$ Pauli operators is
given by
\begin{align}
  \label{eq:ActionSUM}
  \text{SUM}: X_1(q) \otimes I_2 &\to X_1(q) \otimes X_2(q) \, ,
  \nonumber \\
  Z_1(p) \otimes I_2 &\to Z_1(p) \otimes I_2 \, , \nonumber \\
  I_1 \otimes X_2(q) &\to I_1 \otimes X_2(q) \, , \nonumber \\
  I_1 \otimes Z_2(p) &\to Z_1(p)^{-1} \otimes Z_2(p) \, . 
\end{align}
This gate describes the unitary transformation used in a back-action
evasion or quantum nondemolition process~\cite{Wal94}.

The Fourier transform $F$ is the CV analog of the Hadamard
transformation.  It is defined as
\begin{equation}
  \label{eq:DefFourier}
  F = \exp \Bigl( \frac{i}{\hbar} 
  \frac{\pi}{4} (\hat{q}^2 + \hat{p}^2) \Bigr) \, ,
\end{equation}
and the action on the Pauli operators is
\begin{align}
  \label{eq:ActionFourier}
  F: X(q) &\to Z(q) \, ,  \nonumber \\
  Z(p) &\to X(p)^{-1} \, .
\end{align}
The `phase gate' $P(\eta)$ is a squeezing operation for CV, defined by
\begin{equation}
  \label{eq:DefPhase}
  P(\eta) = \exp \Bigl( \frac{i}{2\hbar} \eta \hat{q}^2 \Bigr) \, ,
\end{equation}
and the action on the Pauli operators is
\begin{align}
  \label{eq:ActionPhase}
  P(\eta): X(q) &\to e^{\frac{i}{2\hbar}\eta q^2} 
  X(q)Z(\eta q) \, ,  \nonumber \\
  Z(p) &\to Z(p) \, .
\end{align}
(The operator $P(\eta)$ is called the phase gate, in analogy to the
discrete-variable phase gate $P$~\cite{Got01b}, because of its
similar action on the Pauli operators.)

For discrete variables, it is possible to generate the Clifford group
using only the SUM, $F$, and $P$ gates~\cite{Got01b}.  However, for
the CV definitions above, the operators SUM, $F$, and $P(\eta)$ are
all elements of Sp($2n,\mathbb{R}$); they are generated by homogeneous
quadratic Hamiltonians only.  Thus, they are in a subgroup of the
Clifford group.  In order to generate the entire Clifford group, one
requires a continuous HW($1$) transformation [\textit{i.e.}, a linear
Hamiltonian, that generates a one-parameter subgroup of HW($1$)]
such as the Pauli operator $X(q)$.  This set $\{\text{SUM},\, F,\,
P(\eta),\, X(q); \eta,q \in \mathbb{R} \}$ generates the Clifford group.

We now have the necessary components to prove the main theorem of this
paper regarding efficient classical simulation of a CV process.  We
employ the stabilizer formalism used for discrete variables and follow
the evolution of the Pauli operators rather than the states.  To start
with, let us consider the ideal case of a system with an initial state
in the computational basis of the form $|q_1,\,
q_2,\,\ldots,\,q_n\rangle$. This state may be fully characterized by
the eigenvalues of the generators of $n$ Pauli operators $\{\hat
q_1,\, \hat q_2,\ldots,\,\hat q_n\}$.  Any continuous variable process
or algorithm that is expressed in terms of Clifford group
transformations can then be modeled by following the evolution of the
generators of these $n$ Pauli operators, rather than by following the
evolution of the states in the Hilbert space
$\mathcal{L}^2(\mathbb{R}^n)$.  The Clifford group maps linear
combinations of Pauli operator generators to linear combinations of
Pauli operator generators (each $\hat{q}_i$ and $\hat{p}_i$ is mapped
to sums of $\hat{q}_j$, $\hat{p}_j$, $j=1,\ldots,n$ in the Heisenberg
picture).  For each of the $n$ generators describing the initial
state, one must keep track of $2n$ real coefficients describing this
linear combination.  To simulate such a system, then, requires
following the evolution of $2n^2$ real numbers.

In the simplest case, measurements (in the computational basis) are
performed at the end of the computation. An efficient classical
simulation involves simulating the statistics of linear combinations
of Pauli operator generators.  In terms of the Heisenberg evolution,
the $\hat{q}_j$ are described by their initial eigenvalues, and the
$\hat{p}_j$ in the sum by a uniform random number. This prescription
reproduces the statistics of all multi-mode correlations for
measurements of these operators.

Measurement in the computational basis plus feed-forward \emph{during}
the computation may also be easily simulated for a sufficiently
restricted class of feed-forward operations; in particular, operations
corresponding to feed-forward displacement (not rotation or squeezing,
though this restriction will be dropped below) by an amount
proportional to the measurement result.  Such feed-forward operations
may be simulated by the Hamiltonian that generates the SUM gate with
measurement in the computational basis delayed until the end of the
computation. In other words, feed-forward from measurement can be
treated by employing conditional unitary operations with delayed
measurement~\cite{Nie00}, thus reducing feed-forward to the case
already treated.

In practice, infinitely squeezed input states are not available.
Instead, the initial states will be of the form
\begin{equation}
  \hat S_1(r_1)\otimes\hat S_2(r_2)\otimes\cdots\otimes \hat S_n(r_n)
  |0, 0, \ldots, 0\rangle \,,
\end{equation}
where $|0\rangle$ is a vacuum state and $\hat S(r)$, $r\in \mathbb{R}$
is the squeezing operation which can be expressed directly in terms of
elements of the Clifford group.  Now the vacuum states may \emph{also}
be described by stabilizers $\{ \hat q_1 + i \hat p_1,\, \hat q_2 + i
\hat p_2,\, \ldots,\, \hat q_n + i \hat p_n\}$ which are complex
linear combinations of the generators.  Combining the initial
squeezing operators into the computation, a classical simulation
requires following the evolution of $4n^2$ numbers (twice that of
infinitely squeezed inputs due to the real and imaginary parts).
Measurements in the computational basis are again easily simulated in
terms of this Heisenberg evolution, by treating each of the $q_i$ and
$p_i$ as random numbers independently sampled from a Gaussian
distribution with widths described by the vacuum state.  Simulation of
measurement plus feed-forward follows exactly the same prescription as
before.

The condition for ideal measurements can be relaxed.  Finite
efficiency detection can be modeled by a linear loss mechanism
\cite{Yue80}.  Such a mechanism may be described by quadratic
Hamiltonians and hence simulated by the Clifford group.  Note that the
Clifford group transformations are precisely those that preserve
Gaussian states; \textit{i.e.}, they transform Gaussians to Gaussians; this
observation allows us to remove our earlier restriction on
feed-forward gates and allow for classical feed-forward of any
Clifford group operation.  Note that non-Gaussian components to the
states cannot be modeled in this manner.

Finally, it should be noted that modeling the evolution requires
operations on real-valued (continuous) variables, and thus must be
discretized when the simulation is done on a discrete (as opposed to
analog) classical computer. The discretization assumes a finite error,
which will be bounded by the smaller of the initial squeezing or the
final detector `resolution' due to finite efficiency, and this error
must remain bounded throughout the simulation.  As only the
operations of addition and multiplication are required, the
discretization error can be kept bounded with a polynomial cost
to efficiency.

Thus, we have proved the extension of the GK theorem for
continuous variables:

\textbf{Theorem 2 (Efficient Classical Simulation):} \textit{Any
  continuous variable quantum information process that initiates with
  Gaussian states (products of squeezed displaced vacuum states) and
  performs only (i) linear phase-space displacements (given by the
  Pauli group), (ii) squeezing transformations on a single oscillator
  system, (iii) {\rm SUM} gates, (iv) measurements in position- or
  momentum-eigenstate basis (measurements of Pauli group operators)
  with finite losses, and (v) Clifford group {\rm
    [HW($n$)]Sp($2n,\mathbb{R}$)} operations conditioned on classical
  numbers or measurements of Pauli operators (classical feed-forward),
  can be \emph{efficiently} simulated using a classical computer.}

We could summarize the conditions \textit{(i-iii)} by simply stating
\textit{(i-iii) transformations generated by Hamiltonians that are
  inhomogeneous quadratics in the canonical operators $\{ \hat{q}_i,\,
  \hat{p}_i;\, i=1,\ldots,n \}$}, which is equivalent.  Thus, any
circuit built up of components described by one- or two-mode quadratic
Hamiltonians [such as the set of gates SUM, $F$, $P(\eta)$, and
$X(q)$], that initiates with finitely squeezed states and involves
only measurements of canonical variables may be efficiently
classically simulated.

As with the discrete-variable case, these conditions do not mean that
entanglement between the $n$ oscillator systems is not allowed; for
example, starting with (separable) position eigenstates, the Fourier
transform gate combined with the SUM gate can lead to entanglement.
Thus, algorithms that produce entanglement between systems may still
satisfy the conditions of the theorem and hence may be simulated
efficiently on a classical computer; included are those used for CV
quantum teleportation~\cite{Bra98}, quantum
cryptography~\cite{Ral00,Hil00,Rei00,Got01a}, and error correction for
CV quantum computing~\cite{Bra98b,Bra98c}.  Although these processes
are of a fundamentally quantum nature and involve entanglement between
systems, this theorem demonstrates that they do not provide any
speedup over a classical process.  Thus, our theorem provides a
valuable tool in assessing the classical complexity of simulating
these quantum processes.

As shown in~\cite{Llo99}, in order to generate all unitary
transformations given by an arbitrary polynomial Hamiltonian (as is
necessary to perform universal CV quantum computation), one must
include a gate described by a Hamiltonian other than an inhomogeneous
quadratic in the canonical operators, such as a cubic or higher-order
polynomial.  Transformations generated by these Hamiltonians do not
preserve the Pauli group, and thus cannot be described by the
stabilizer formalism.  Moreover, \emph{any} such Hamiltonian is
sufficient~\cite{Llo99}.  One example would be to include an optical
Kerr nonlinearity~\cite{Mil83}, but there is a lack of sufficiently
strong nonlinear materials with low absorption.  Alternatively, it has
recently been proposed that a measurement-induced nonlinearity (using
ideal photodetection) could be used in an optical scheme without the
need for nonlinear materials in the computation~\cite{Got01b,Bar01}.
The physical realization of such nonlinearities is an important quest
for quantum information theory over continuous variables.  These
nonlinear transformations can be used in CV algorithms that do not
satisfy the criteria of this theorem, and which may provide a
significant speedup over any classical process.

\begin{acknowledgments}
  This project has been supported by an Australian Research Council
  Large Grant.  SDB acknowledges the support of a Macquarie University
  Research Fellowship.  SLB and KN are funded in part under project
  QUICOV as part of the IST-FET-QJPC programme.
\end{acknowledgments}

\end{document}